       \newcommand{\beq}{\begin{equation}}
       \newcommand{\eeq}{\end{equation}}
       \newcommand{\beqa}{\begin{eqnarray}}
       \newcommand{\eeqa}{\end{eqnarray}}
       \newcommand{\beqas}{\begin{eqnarray*}}
       \newcommand{\eeqas}{\end{eqnarray*}}
       
       \newcommand{\bnab}{\mbox{\boldmath ${\nabla}$}}

       \newcommand{\x}{\mbox{\boldmath$\times$}}

       \newcommand{\be}{{\mathbf e}}
       \newcommand{\bj}{{\mathbf j}}

       \newcommand{\bv}{{\mathbf v}}

       \newcommand{\bB}{{\mathbf B}}

       \newcommand{\bE}{{\mathbf E}}

       \newcommand{\dadb}[2]{\frac{{  d}#1}{{  d}#2}}

\newcommand{\nonu}{\nonumber \\}

\documentclass[12pt]{article}
\usepackage[dvips]{graphicx}\usepackage[dvips]{graphics}
\usepackage{enumerate}
\usepackage{psfrag}
\usepackage{color}
\begin{document}
\begin{center}
{\large \bf
%On equilibria in Hall magnetohydrodynamics model
 Magnetohydrodynamic ``cat eyes" and stabilizing effects of plasma
 flow}
\vspace{3mm}

{\large  G. N. Throumoulopoulos$^1$, H. Tasso$^2$, G.
Poulipoulis$^1$}
\vspace{2mm}

{\it
$^1$University of Ioannina, Association Euratom - Hellenic Republic,\\
%\vspace{-1mm}
 Section of Theoretical Physics, GR 451 10 Ioannina, Greece}
\vspace{2mm}

 { \it  $^2$Max-Planck-Institut f\"{u}r Plasmaphysik, Euratom
Association,\\
%\vspace{-1mm}
 D-85748 Garching, Germany }
\end{center}
%\noindent
%
%
%\vspace{2mm}
%\begin{center}
%{\large \it December 2000}
%\end{center}
\vspace{2mm}
\begin{center}
{\bf Abstract}
\end{center}
 \noindent

 The cat-eyes  steady state solution in the framework of
 hydrodynamics describing an infinite row of
 identical vortices is extended to the magnetohydrodynamic (MHD)
equilibrium equation with incompressible flow of arbitrary
direction. The extended solution covers a variety of equilibria
including one- and two-dimensional generalized force-free and
Harris-sheet configurations which are preferable  from those usually
employed
%in the literature
as initial states in reconnection studies. Although the vortex shape
is not affected by the magnetic field, the flow  in conjunction with
the equilibrium nonlinearity has a strong impact on isobaric
surfaces by forming pressure islands located within the cat-eyes
vortices. More importantly,  a magnetic-field-aligned flow of
experimental fusion relevance and the flow shear have significant
stabilizing effects in the region of pressure islands. The stable
region is enhanced by an external axial (``toroidal") magnetic
field.

\newpage
\begin{center}
{\bf \large I.\  Introduction}
\end{center}

Sheared flows   influence the equilibrium and stability properties
of magnetically confined plasmas and result in  transitions to
improved modes either in the edge region (low-to-high-mode
transition) or in the central region (internal transport barriers)
of fusion devices.
%, e.g. Ref. \cite{Te}.
 As concerns equilibrium, the
convective velocity term in the momentum equation makes the isobaric
surfaces to deviate from magnetic surfaces,
 unlike the case of quasistatic steady states \cite{qua},
 thus potentially affecting stability. For
symmetric two dimensional equilibria, this effect has been examined
on the basis of analytic solutions to linearized forms of
generalized Grad-Shafranov equations, e.g. Ref. \cite{MaPe}. For
flows of fusion concern, i.e. for Alfv\'en Mach numbers of the order
of 0.01,  this deviation is small  and consequently isobaric and
magnetic surfaces have the same topology.

\begin{figure}[!h]
 \begin{center}
 \includegraphics[scale=0.8]{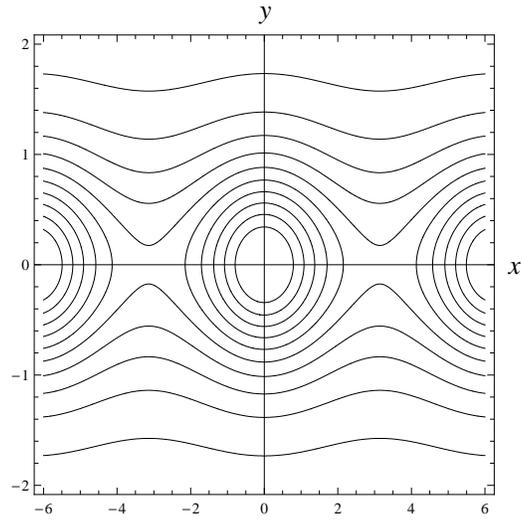}
 \caption{$u$-lines of the MHD cat-eyes solution (\ref{7}) for $\epsilon=0.2$
  and $\beta=0.02$ as intersections of the magnetic surfaces with the
                     poloidal plane.}
 \label{fig:1a}
 \end{center}
 \end{figure}

Aim of the present study is to examine the impact of  flow in
conjunction with nonlinearity to certain equilibrium and stability
properties in relation to the departure of the isobaric from
magnetic surfaces. Motivation was
 a solution of a nonlinear form of the
hydrodynamic equation describing the steady motion of an inviscid
incompressible  fluid in two dimensional plane geometry, known as
``cat eyes", which represents an infinite row of identical vortices
(\cite{Stu,Pe}; see also Fig. \ref{fig:1a}). This solution is
extended here to the MHD equilibrium equation  with incompressible
flow. Then,  the stability of the extended solution is examined by
means of a recent sufficient condition \cite{ThTa1}. The major
conclusion is that owing to the nonlinearity of the equilibrium, the
flow and flow shear affect drastically the pressure surfaces and
have significant stabilizing effects in the region of modified
pressure.

The MHD equilibrium equations with incompressible flow for
translationally symmetric plasmas  are reviewed  in Sec. II. In Sec.
III a solution of the pertinent generalized Grad-Shafranov equation
describing a whole set of equilibria is constructed as an extension
of the cat-eyes solution. Then for parallel flows and constant
density the stability of the solution obtained  is studied in Sec.
IV. Sec. V recapitulates the study and summarizes the conclusions.

\begin{center}
 {\bf \large II.\ \ Review of the equilibrium equations}
 \end{center}

 The MHD equilibrium states of a translationally  symmetric  magnetized plasma
 with incompressible flows
 satisfy the generalized Grad-Shafranov equation
  \cite{ThTa2},
\beq (1-M^2) \nabla^2\psi -
         \frac{1}{2}(M^2)^\prime |\nabla \psi|^2
                     + \left(Ps+\frac{B_z^2}{2}\right)^\prime=0
                            \label{1}
 \eeq
 and the Bernoulli relation for the pressure
 \beq
 P=P_s(\psi) - \frac{1}{2}M^2\left|\nabla \psi\right|^2.
                          \label{2}
 \eeq
 Here,  the function
 $\psi(x,y)$
 %$\psi(\xi, \eta)$
 labels the magnetic surfaces  with   $(x,y,
 z)$ Cartesian coordinates
 so that $z$ corresponds to the axis of symmetry and ($x, y$) are associated with
 the poloidal plane; $M(\psi)$ is
 the Mach function of the poloidal velocity with respect to the
 poloidal-magnetic-field Alfv\'en velocity;
  $B_z$ is the axial
  %(``toroidal")
   magnetic field;
 for vanishing flow the surface function $P_s(\psi)$
  coincides with the pressure; the prime denotes a derivative with respect to $\psi$.
  The surface quantities
$M(\psi)$,  $B_z(\psi)$ and $P_s(\psi)$ are free functions for
each choice of which
 (\ref{1})  is fully determined and can be solved whence the boundary
condition for $\psi$ is given. Also, to completely determine the
equilibrium, a couple of additional   surface functions are needed,
i.e, the density, $\varrho(\psi)$, and the electrostatic potential,
$\Phi(\psi)$. Details including derivation of  (\ref{1}) and
(\ref{2}) can be found  in Refs. \cite{ThTa2,SiTh}.

Eq. (\ref{1}) can be  simplified  by  the  transformation
\cite{Cl,Mo}
\begin{equation}
u(\psi) = \int_{0}^{\psi}\left\lbrack 1 -
M^{2}(g)\right\rbrack^{1/2} dg,
                                    \label{2a}
\end{equation}
which reduces (\ref{1})  to
\beq
  \nabla^2 u+\frac{d}{du}\left(P_s+\frac{B_z^2}{2}\right)=0.
                                \label{3}
\eeq
 %%
%Here the primes indicate now derivatives with respect to $u$
%$F^\prime= d F/d\psi$  and$\Phi^\prime= d \Phi/d\psi$.
Also, (\ref{2}) is put in the form
 \beq
 P=P_s(u) - \frac{M^2}{2(1-M^2)}\left|\bnab u\right|^2.
                          \label{4}
 \eeq
 Note that (\ref{3}) free of  a quadratic term as $|{\bf\nabla}u|^{2}$ is identical in form with
 the quasistatic MHD equilibrium equation as well as to the equation governing the steady motion of an inviscid
 incompressible fluid in the framework of hydrodynamics. Transformation (\ref{2a}) does not affect the
 magnetic surfaces, it just relabels them.  Also, once a solution of (\ref{3}) is
 found,
 the equilibrium can be completely constructed in the
 $u$-space; in particular, the magnetic field, current density,
 velocity, and electric field can be determined  by the relations:
 \beqa
 \bB&=&B_z\be_z+\left(1-M^2\right)^{-1/2}\be_z\x\bnab u \\
 \bj&=&\left\lbrack \left(1-M^2\right)^{-1/2}\bnab^2 u
 +\frac{1}{2}\dadb{M^2}{u}\left(1-M^2\right)^{-3/2}\left| \bnab
 u\right|^2\right\rbrack \be_z \\ \nonumber
 && -\dadb{B_z}{u}\be_z\x\bnab u \\
 \bv&=&\frac{M}{\sqrt{\varrho}}\bB-\left(1-M^2\right)^{-1/2}\dadb{\Phi}{u}\be_z
 \\
 \bE&=&-\dadb{\Phi}{u}\bnab u .
 \eeqa

 Analytic solutions to linearized forms of
 (\ref{3}) have been constructed  for quasistatic \cite{Ga,Ba} and stationary equilibria \cite{SiTh}.
 As already mentioned in Sec. I, for flows of experimental fusion
 relevance ($|M|\approx 0.01$) the  departure of the
 isobaric from magnetic surfaces is small (see for example Fig. 2  of Ref.
 \cite{MaPe}),
 so that
 the topology of these two families of surfaces is identical.

 \begin{center}
 {\bf \large II.\ \ Magnetohydrodynamic ``cat eyes" with flow}
 \end{center}

 %In the framework of hydrodynamics
 %a nonlinear solution in plane geometry known as ``cat eyes" consists of an infinite row of identical vortices
 %\cite{Pe,Stu}.
 The present section aims at extending the hydrodynamic cat-eyes solution  to
   (\ref{3})
 and examine certain equilibrium characteristics in connection with the   impact of the flow together with
 nonlinearity.
 For convenience we introduce dimensional quantities: $\tilde{x}=x/L$,
 $\tilde y=y/L$, $\tilde{u}=u/(B_{z0}L)$,  $\tilde{\varrho}=\varrho/\varrho_0$,
  $\tilde{P}=P/B_{z0}^2$, $\tilde{\bB}=\bB/B_{z0}$, $\tilde{\bj}=\bj/(B_{z0}/L)$,
 $\tilde{\bv}=\bv/v_{A0}$, where $v_{A0}=B_{z0}/\sqrt{\varrho_0}$ ,
 and $\tilde{\bE}=\bE/(B_{z0} v_{A0})$; here,
   $L$, $B_{z0}$, and
 $\varrho_0$ are  reference quantities to be defined later.
   Eqs. (\ref{3}) and (\ref{4}) hold in identical forms for the tilted
   quantities and
    will be further employed  as  dimensionless  by dropping
   for simplicity the tilde.   To construct a
    cat-eyes  solution we make the ansatz
    \beq
    \dadb{(P_s+B_z^2/2)}{u} = (\epsilon^2-1) \exp(-2u),
                                                  \label{5}
    \eeq
  by which (\ref{3}) reduces to the following form of
  Liouville' s equation:
  \beq
  \nabla^2 u=(1-\epsilon^2)\exp(-2 u).
                                           \label{6}
  \eeq
  Eq. (\ref{6}) admits the solution
  \beq
  u=\ln\left\lbrack \cosh (y) -\epsilon \cos (x)\right\rbrack,
                                         \label{7}
  \eeq
  the characteristic lines  of which are shown in Fig. \ref{fig:1a}.
  The parameter $\epsilon$
  determines the vortex size;  for $\epsilon=1$ the solution represents an infinite row of
  point
  vortices and for $\epsilon=0$  it becomes one-dimensional:
  $
  u=\ln\cosh y
  $.
  It is noted here that though (\ref{7}) is
  singular in the limit of $y \rightarrow \infty$, all the local equilibrium quantities are everywhere regular.
  Eq. (\ref{5}) can be solved for $P_s(u)+B_z^2/2$ to
  yield
  \beq
  P_s+\frac{B_z^2}{2}=\frac{1-\epsilon^2}{2}\exp{(-2
  u)}+c_0=\frac{1-\epsilon^2}{2\left(\cosh y -\epsilon\cos
  x\right)^2}+c_0,
                                        \label{9}
  \eeq
  where $c_0$ is a constant.
 The equilibria described by (\ref{7}) and (\ref{9})  have the following
  characteristics:
  \begin{enumerate}
  \item  The vortices are by construction of solution (\ref{7}) identical to the
  respective hydrodynamic vortices, viz., the  magnetic field does not affect the vortex  shape.
 \item  Since  magnetic field and  current density lie  on  the velocity or magnetic surfaces, the vortices
  can be regarded as magnetic islands with plasma flow. Quasistatic
   MHD and hydrodynamic cat eyes can be recovered as
  particular cases. Also, it may be noted that for flows non parallel to
  the magnetic field,  the electric field is perpendicular to the magnetic
  surfaces [Eq. (9)].
   \item In fact, (\ref{7}) and (\ref{9}) hold for a rather large set of equilibria because the
     functions $\rho(u)$, $\Phi(u)$, $M(u)$ and one out of $B_z(u)$ and $P_s(u)$ remain free.
 \end{enumerate}

  We will further consider a subset of steady sates by assigning the
  free functions $P_s$, $B_z$ and $M$ as
  \beq
  P_s(u)=\beta\frac{1-\epsilon^2}{2}\exp{(-2u)} + \frac{\beta_f}{2},
                                                 \label{9a}
  \eeq
  \beq
  B_z^2(u)=(1-\beta)(1-\epsilon^2)\exp{(-2u)}+B_{z0}^2,
                                                 \label{10}
  \eeq
   \beq
  M=M_0\exp(-2nu)=M_0\left(\cosh y-\epsilon \cos x\right)^{-2n},\ \ n>0.
                                                     \label{11}
  \eeq
 Choice (\ref{11}) yields a peaked $M^2$-profile along $y$ with $|M_0|$ being the maximum absolute value
 at $x=y=0$. The profile becomes steeper as $n$ takes larger positive
 values, thus increasing the shear of $M$ in relation to the velocity shear.
 Henceforth, profiles will refer to the $y$-axis. The parameter $B_{z0}$  represents the
 external axial magnetic field,
 $$
 \beta=\frac{P_s(\epsilon=y=0)}{B_{z0}^2/2}
 $$
 and
  $\beta_f=P_{s0}/(B_{z0}^2/2)$, where $P_{s0}=\mbox{const}$.
  Note that $\beta$ has been introduced  in (\ref{9a}) and (\ref{10}) in such a way
  that (\ref{9}) is automatically satisfied.
  The other parameter $\beta_f$ in
  (\ref{9a})
   yields force-free quasistatic
  equilibria when $\beta=0$. For $\beta\neq 0$,  we set $\beta_f=0$
  in order that  $P_s $ vanishes for $y \rightarrow \infty $;
  thus, only one of the parameters $\beta$ and $\beta_f$ is finite in
   connection with peaked and flat $P_s$-profiles, respectively. For flat
   $P_s$-profiles,  to guarantee positiveness of the pressure for
   $\beta_f\geq
  0$, Eq. (\ref{4}) is  modified to:
  \beq
  P=\frac{\beta_f}{2} - \frac{M^2}{2(1-M^2)}\left|\bnab u\right|^2
  +\frac{M_0^2}{2(1-M_0^2)}.
                          \label{12}
  \eeq
  The parameters $M_0$,
  and $n$ are free together with $L$, $\varrho_0$, $\epsilon$, $B_{z0}$, and $\beta$ or $\beta_f$.
   It is recalled that dimensionless quantities
  are employed and therefore $B_{z0}=1$. Also, the reference quantities $L$ and $\varrho_0$
  not appearing explicitly in the  equations can  arbitrarily be defined as the
  vortex length (along the $x$-axis) and the density at $x=y=0$.
 Because of the many free parameters,
  there is a variety of steady states  including extensions of
  equilibria  employed as initial states in reconnection studies (see for example Ref. \cite{Ka}).
  An example concerns the one-dimensional, force free quasistatic equilibrium
  recovered for  $\beta=\epsilon=0$. In the presence of flow
    and $\epsilon\neq 0$ this  equilibrium becomes two-dimensional with hollow  pressure
  profile  (Eq. (\ref{12}); see also Fig. 4b). Also, in this case both current density and velocity
  have all three components finite. Another example for $\beta=1$ is an
   equilibrium with $B_z=B_{z0}$,  axial current density and three-component
   velocity.
 For vanishing flow and $\epsilon=0$  this  reduces to the Harris
  sheet equilibrium \cite{Ha}.

 We have examined  the pressure by Mathematica 6  within broad regions of the free parameters,
 i.e.,
 $0\leq\epsilon\leq 1$, $0\leq \beta  \leq 0.9$, $0\leq M_0\leq 0.9$ and $0\leq n \leq 15$.
 Note that, because of the flow term in (\ref{4}) the pressure for certain parametric values can become negative.
 Thus, particular care has been taken in getting  everywhere physically acceptable pressure.
 For two dimensional equilibria, it turns out that the flow has strong impact on the isobaric surfaces by
 creating ``pressure islands" within the cat eyes. This is
 shown in Fig. \ref{fig:2a}. Also $P$-profiles are presented
 in Fig. \ref{fig:3a}. As  can be seen in Fig. 3
 pressure islands appear even  for parametric values  of experimental
 fusion concern ($\beta=0.02$, $M_0=0.02$). Since for linear equilibria the
 flow impact on the pressure is weak, it is the
  nonlinearity here which should play an important role.
  Also, as will be discussed  in Sec. III,
 the formation of pressure islands   may be related  to appreciable stabilizing effects of the flow.

\begin{center}
 {\bf \large II.\ \ Stabilizing effects of the flow}
 \end{center}

 The  stability of the equilibria described by (\ref{7}) and
 (\ref{9a})-(\ref{11})
is now examined by applying  a recent sufficient condition
\cite{ThTa1}. This condition states that a general steady state of a
plasma of constant density and incompressible flow parallel to $\bB$
is linearly stable to small three-dimensional perturbations if  the
flow is sub-Alfv\'enic ($M^2<1$) and $A\geq 0$, where $A$ is given
by Eq. (20) of Ref. \cite{ThTa1}. Consequently, we restrict the
study to parallel flows and set $\varrho=1$. First it is noted that
on the basis of Mercier expansions it turns out that the condition
is never satisfied in the vicinity of the magnetic axis ($A< 0$)
\cite{ind}. This holds for generic two-dimensional equilibria
irrespective of the geometry. Also,  for the pressure (\ref{12}),
%, associated with
%force-free quasistatic equilibria,
the quantity $A$
 is independent of $\beta_f$, as may be expected on physical grounds,
  because $A$ contains $dP_s/du$ and not
 $P_s$ itself. In the $u$-space for translationally symmetric
 equilibria,
 $A$ assumes the form
 \beqa
 A&=&-g^2\left\{\left(\bj\x\bnab
 u\right)\cdot\left(\bB\cdot\bnab\right)\bnab
 u+\frac{1}{2}\frac{dM^2}{du}\left(1-M^2\right)^{-1}\left|\bnab
 u\right|^2 \right. \nonu
 & & \left.\left\lbrack \left(1-M^2\right)^{-1/2}\bnab u\cdot\frac{\bnab
 B^2}{2}+g(1-M^2)^{-1}\left|\bnab u\right|^2\right\rbrack \right\},
                                                    \label{12a}
 \eeqa
 where
 \beq
 g=(1-M^2)^{-1/2}\left(\frac{dP_s}{du}-\frac{dM^2}{du}\frac{B^2}{2}\right),
 \eeq
 and $\bB$ and $\bj$ as given by (6) and (7).
 To calculate  $A$
analytically for the equilibria under consideration we developed a
code in Mathematica 6. The expressions obtained for both peaked and
flat $P_s$-profiles being lengthy are not given explicitly here
except for the case of quasistatic equilibria [Eq. (\ref{12b})
below]. The calculations led to the following conclusions.
\begin{enumerate}
\item For quasistatic equilibria ($M_0=0$) the quantity $A$ assumes the concise form
\beq A=\frac{\epsilon(1-\epsilon^2)\left\lbrack
\epsilon\cosh(y)\sin(x)^2
+\cos(x)\sinh(y)^2\right\rbrack}{\left\lbrack\epsilon\cos(x)-\cosh(y)\right\rbrack^5}.
                                                   \label{12b}
\eeq
 Note that $A$ becomes independent of $\beta$ and $B_{z0}$.
 The condition is nowhere satisfied in the island region except for one dimensional
 configurations ($\epsilon=0$), point vortices ($\epsilon=1$), the magnetic axes,  the $x$-points and for
$y\rightarrow \infty$ for which $A=0$. A profile of $A$
%on the mid-planes $x=0$ and $y=0$
is given in Fig. \ref{fig:4}.
 \item The flow  results in the
formation of a stable region close to the magnetic axis  in the
location of pressure islands. An example shown the sign of $A$ on
the poloidal plane is presented in Fig. 6a.
 The red colored regions are stable ($A\geq 0$),  while in the blue colored
 region it holds $A<0$.
 The  whole area of Fig. 6a becomes blue colored when $M_0=0$.
 \item The stable region broadens when the parameters $M_0$, $n$ and $\epsilon$ take larger values
 as can be seen in Figs. 7a, 7b and 7c, respectively. Note the sensitiveness of
 $A$ in the region of the stable window to the small variation  of these
 parameters possibly
   related to the nonlinearity;  in particular, $\epsilon$
 appears in the argument of  the cat-eyes solution (\ref{7}). These results hold for both
 peaked- and flat-$P_s$ equilibrium profiles. Unlikely, the stable region
 is  rather insensitive to the variation of $\beta$. An example is given in Fig. 7d, where the stable window
   persists (just
getting slightly smaller)  when $\beta$ is  increased by an order of
 magnitude  (from 0.02 to 0.2). Also, for point vortices ($\epsilon=1$) $A$ becomes independent of
 $\beta$   irrespective of the value of $M_0$.
 \item Although for $M_0=0$ the vacuum magnetic field $B_{z0}$ has no impact on $A$
 [Eq. (\ref{12b})], in combination with the flow,  $B_{z0}$ can enhance the stable region. An example
 of this synergetic effect is shown in Fig. 8a. Another example of such a strong synergism
 can be seen in Fig. 8b for a two-dimensional Harris-type equilibrium ($\beta=1$, $\epsilon \neq 0$). In this case,
 while the  flow itself can not  make $A$ positive, together with
  $B_{z0}$  it results in the formation of the stable
 window.
 \end{enumerate}

\begin{center}
 {\bf \large V.\ \ Summary and Conclusions}
 \end{center}

 We have extended the  ``cat-eyes" solution of the hydrodynamic equilibrium equation to
 cover  MHD magnetically confined plasmas
  with incompressible flow.
  %The  solution satisfies a  Liouville's type equation and represents
 % an infinite
   %row of identical vortices in plane geometry.
 The extension was accomplished  smoothly because the pertinent generalized Grad-Shafranov equation can be transformed
 to  a form identical
 with that of the  hydrodynamic equation [Eq. (\ref{3})]. Velocity, magnetic field and current density
 of the extended equilibrium share the same
  surfaces; therefore, the vortices  can be viewed as magnetic islands with
  flow with  the magnetic field not affecting the vortex
  shape.
   Also, to be compatible with the cat-eyes solution,
 the axial magnetic field, $B_z(u)$,
 and the quasistatic pressure,
  $P_s(u)$,
  %whith the function $u$ labeling the magnetic surfaces,
  must satisfy relation (\ref{9}).
  The equilibrium is   generic enough because
  four surface quantities, i.e. the density, the electrostatic potential,
 the poloidal Alfv\'en Mach function [$M(u)$] and either
 $B_z(u)$ or $P_s(u)$
  remain free.  Generalized Harris or force free-type  equilibria can be derived as particular cases.
  % owing to
  %the flow they are preferable to those employed usually in the literature as initial states of reconnection studies.
  Furthermore, the flow caused departure of the pressure surfaces from the
  magnetic surfaces  has been examined by assigning the functions $B_z(u)$, $P_s(u)$ and $M(u)$
  [Eqs. (\ref{9a}-\ref{11})].
  The equilibrium   has the following seven free
  parameters:  the island
  length ($L$),  the density $\varrho_0$ on the island axis, the external axial  magnetic field   ($B_{z0}$),
    a parameter $\epsilon$ determining  the island size,
     a local  ratio of the thermal pressure to the
    magnetic pressure ($\beta$ or $\beta_f$ in connection with peaked and flat profiles of $P_s$,
    respectively),
    the Mach number $M_0$ on the island axis, and a
    velocity-shear-related
    parameter $n$. It turns out
 that, unlike to linear equilibria,  the flow strongly affects the pressure surface topology
  by forming   pressure islands on the poloidal plane within the cat eyes, even  for flows of laboratory fusion concern.
 %Since for terestrial-fusion-plasma relevant solutions of linearized
 %forms of the Grad-Shafranov equation
  %the isobaric surfaces
  %deviate only slightly from the magnetic surfaces,
  %this strong modification of pressure surfaces is related to nonlinearity.

  For parallel flows and constant density, the linear stability of the equilibria constructed has been examined by means
  of a recent sufficient condition guaranteeing stability when the flow is sub-Alfv\'enic and an equilibrium dependent
  quantity A [Eq. (\ref{12a})] is nonnegative.
 By symbolic computation of   $A$
 for a broad variation of the parameters
  $\epsilon$, $\beta$, $M_0$ and $n$, we came to the following
  conclusions.  The flow  can result in the formation of a stable region,
  close to the magnetic axis in the location of    pressure islands,
  thus indicating a correlation of stabilization with nonlinearity.
  The stable region can appear for fusion
  relevant values of $ M_0$  on the order of 0.01  when the velocity shear becomes
  appropriately large ($n\approx 10$), enhances as $n$ becomes larger and persists for a large
  variation of $\beta$ (from 0.02 to 0.2).
  Also, the broader the stable region  the larger is the   island size (larger
  $\epsilon$).
  %the velocity- and
  %$B_{z0}$-
  %stabilization  stronger.
  A combination of  velocity and $B_{z0}$ can have synergetic stabilizing effects by
  enlarging the stable region.

  In conclusion, the present study has shown significant stabilizing effects of the flow and flow shear in
  connection with nonlinearity and formation of equilibrium  pressure
  islands. The  study can be extended to several
  directions.
   Firstly, since  four  surface
  functions  remain free in equilibrium  together with many free parameters, there may be a possibility of stability
  optimization.   Secondly,  the problem could be examined   in cylindrical and axisymmetric
  geometries
  in connection with the magnetic field
  curvature and toroidicity.
   Note that in the presence of toroidicity
   non parallel  flows have a stronger impact on equilibrium because,  in addition to the
   pressure,
   they result in a deviation of
  the current density
  surfaces  from the magnetic surfaces. Although in non-plane geometries    nonlinear solutions
  in general should   be constructed numerically, it is interesting
  to pursue
   analytic  translationally symmetric solutions in cylindrical geometry as a next step to the cat-eyes solution.
   At last, it is recalled that the search for necessary and
   sufficient stability conditions with flow remains a tough problem
   as already known in the framework of hydrodynamics.
  %the sufficient condition employed here is rather exceptional because

 \begin{center}
 {\large\bf Acknowledgements }
 \end{center}

 %The authors would like to thank the anonymous referee for critical comments,
 %and particularly for his statement on  an  extension of
 %the local proof  which  resulted in an
 %improved version of the work.

 Part of this work was conducted during a visit of  the first and third authors
 to the Max-Planck-Institut f\"{u}r Plasmaphysik, Garching.
 The hospitality of that Institute is greatly appreciated.

 This work was performed  within the participation of the University
of Ioannina in the Association Euratom-Hellenic Republic, which is
supported in part by the European Union and by the General
Secretariat of Research and Technology of Greece. The views and
opinions expressed herein do not necessarily reflect those of the
European Commission.
 \newpage

  \newpage
\vspace*{-1cm}
 \begin{center}
 {\large \bf  Figure captions}
 \end{center}

 \noindent
 Fig. \ref{fig:1a}: \ $u$-lines of the MHD cat-eyes solution (\ref{7}) for
 $\epsilon=0.2$ and $\beta=0.02$
 as intersections of the magnetic surfaces with the
                     poloidal plane.
 \vspace{0.3cm}

\begin{figure}[!h]
 \begin{center}
 \includegraphics[scale=0.8]{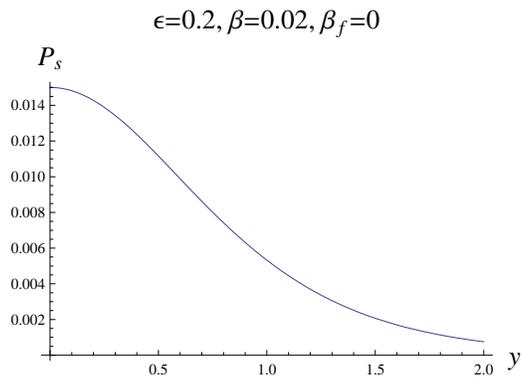}
 \caption{Profile of the quasistatic pressure function $P_s$ [Eq. (\ref{9a})] along the $y-$axis.}
 \label{fig:1b}
 \end{center}
 \end{figure}

\begin{figure}[!h]
 \begin{center}
 \includegraphics[scale=0.8]{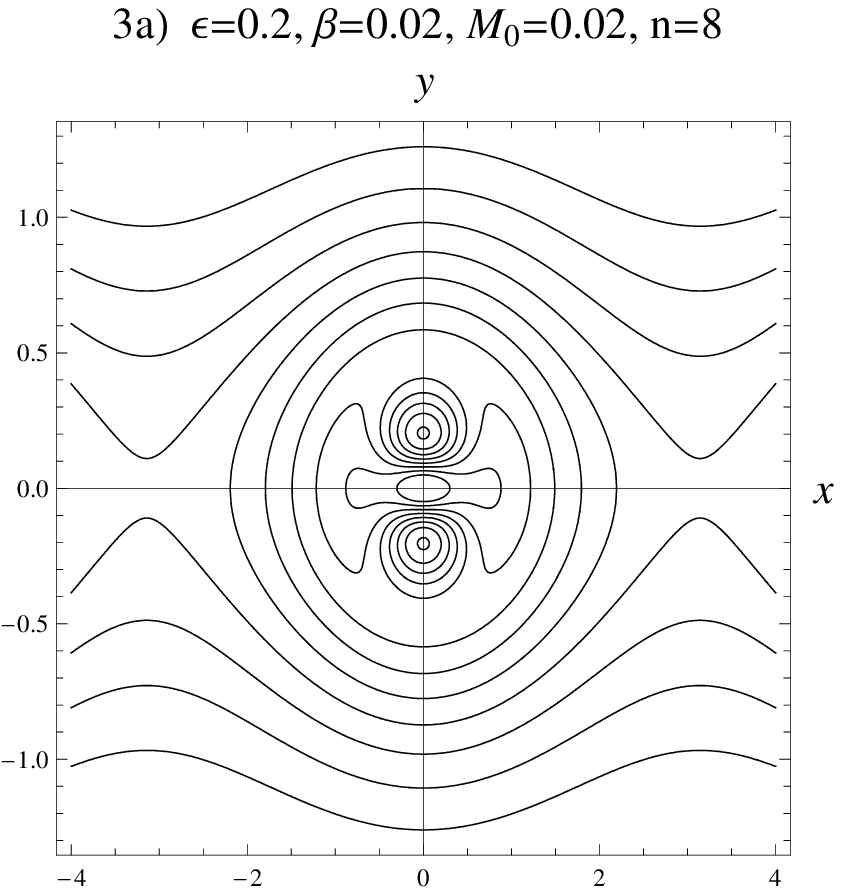}
  \includegraphics[scale=0.8]{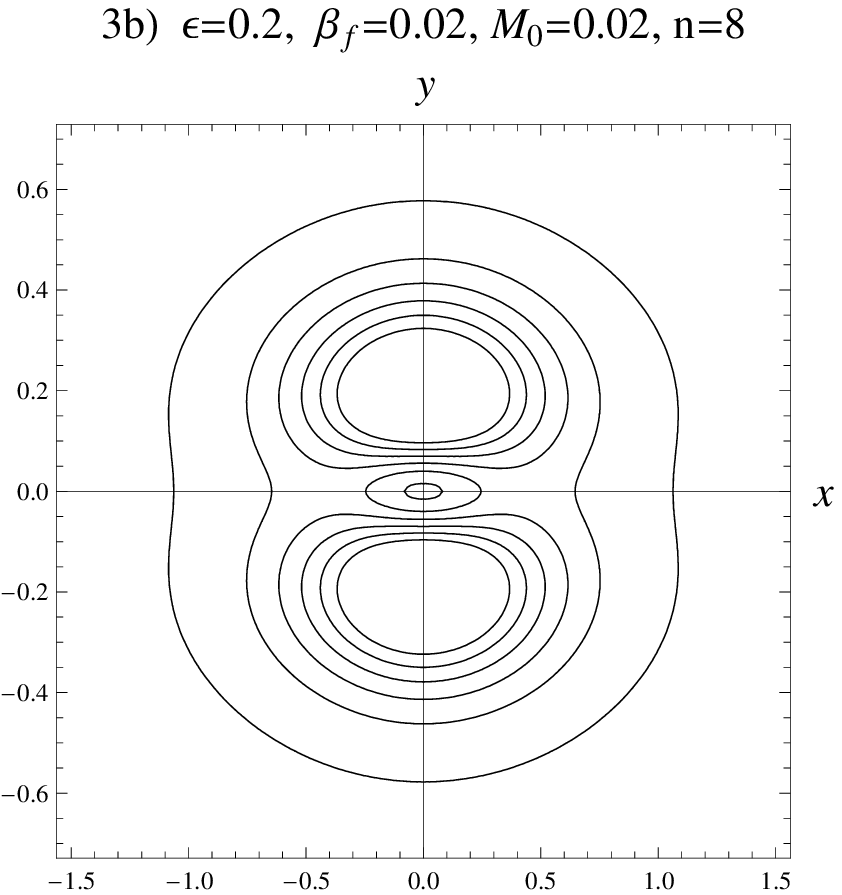}
 \caption{ Pressure islands in connection with Eqs. (\ref{4}) [Fig.
 3a] and (\ref{12}) [Fig. 3b].
 The curves represent pressure lines on the poloidal
 plane.  In the absence of flow the lines of Fig. 3a  coincide with the $u$-lines of
  Fig. \ref{fig:1a} while the equilibrium of Fig. 3b becomes
  force-free.}
 \label{fig:2a}
 \end{center}
 \end{figure}

\begin{figure}[!h]
 \begin{center}
 \includegraphics[scale=0.8]{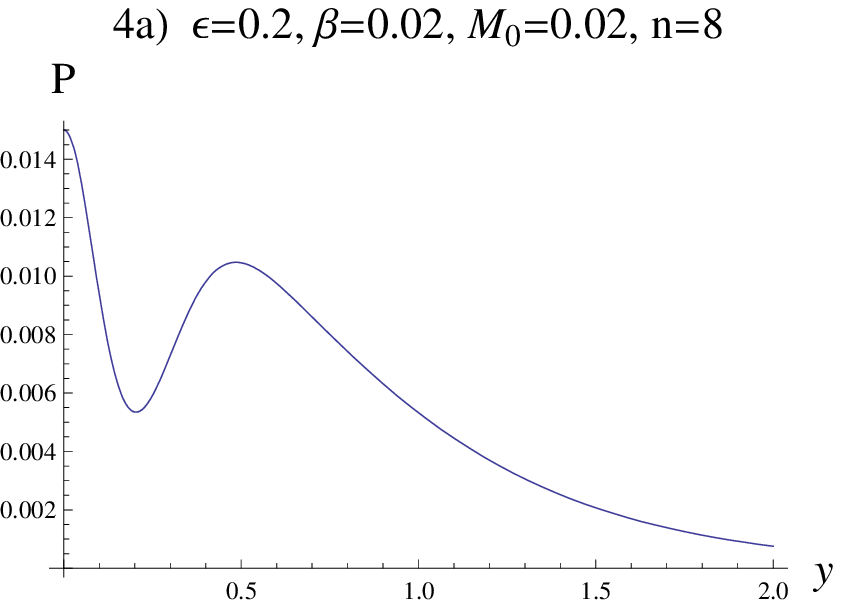}
  \includegraphics[scale=0.8]{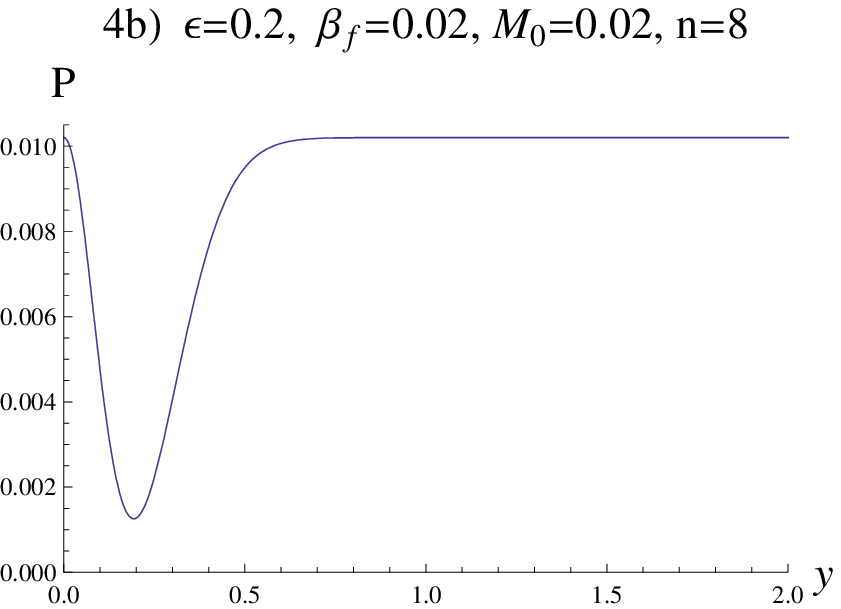}
 \caption{ Pressure profiles along the $y$-axis respective to the
 pressure-island configurations
 3a and 3b. For vanishing flow the profiles 4a and 4b become peaked and   flat, respectively.}
 \label{fig:3a}
 \end{center}
 \end{figure}

\begin{figure}[!h]
 \begin{center}
 \includegraphics[scale=0.8]{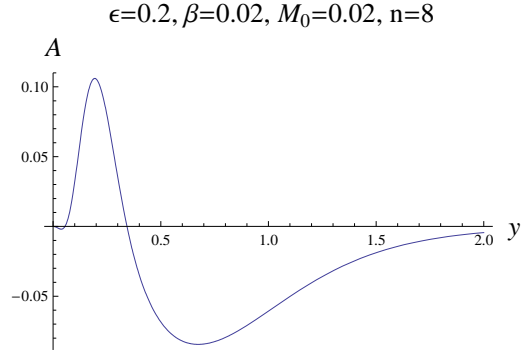}
\caption{Profile of the quantity $A$ [Eq. (\ref{12a})] associated
         with the sufficient condition for linear stability
            for a quasistatic equilibrium ($M_0=0$). Except for the marginally stable points $y=0$ and
            $y\rightarrow \infty$ the condition is nowhere else
            satisfied.}
 \label{fig:4}
 \end{center}
 \end{figure}

\noindent
 Fig. \ref{fig:1b}: \ Profile of the quasistatic pressure function $P_s$ [Eq. (\ref{9a})] along the $y-$axis.
 \vspace{0.3cm}

\noindent
 Fig. \ref{fig:2a}: \  Pressure islands in connection with Eqs. (\ref{4}) [Fig.
 3a] and (\ref{12}) [Fig. 3b].
 The curves represent pressure lines on the poloidal
 plane.  In the absence of flow the lines of Fig. 3a  coincide with the $u$-lines of
  Fig. \ref{fig:1a} while the equilibrium of Fig. 3b becomes
  force-free.
 \vspace{0.3cm}

 \noindent
 Fig. \ref{fig:3a}: \ Pressure profiles along the $y$-axis respective to the
 pressure-island configurations
 3a and 3b. For vanishing flow the profiles 4a and 4b become peaked and   flat, respectively.
 \vspace{0.3cm}

 \noindent
 Fig. \ref{fig:4}: \ Profile of the quantity $A$ [Eq. (\ref{12a})] associated
         with the sufficient condition for linear stability
            for a quasistatic equilibrium ($M_0=0$). Except for the marginally stable points $y=0$ and
            $y\rightarrow \infty$ the condition is nowhere else
            satisfied.
 \vspace{0.3cm}

\begin{figure}[!h]
 \begin{center}
 \includegraphics[scale=0.8]{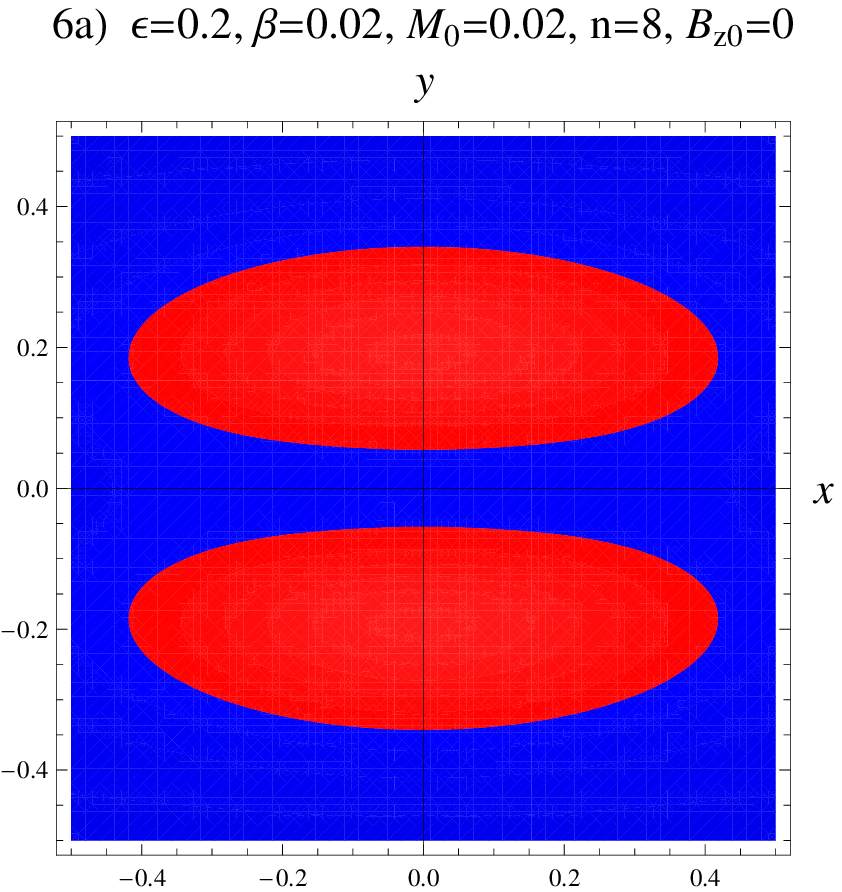}
 \includegraphics[scale=0.8]{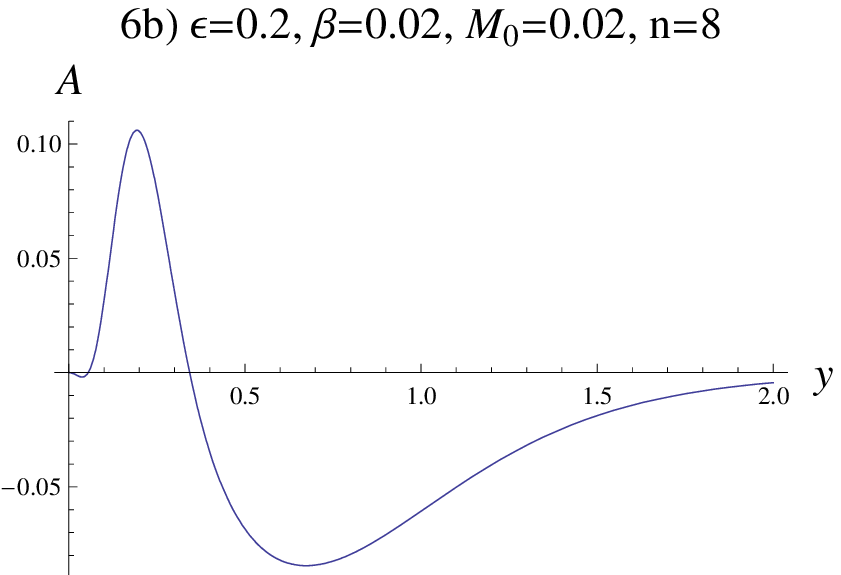}
 \caption{{\em Stabilization effect of flow}: In the presence of flow
 the red colored stable
 regions
 appear
  in the diagram 6a where $A\geq 0$. The respective stable window can be seen in the profile of $A$ in 6b.}
 \label{fig:6a}
 \end{center}
 \end{figure}

 \noindent
 Fig. \ref{fig:6a}: \ {\em Stabilization effect of flow}: In the presence of flow
 the red colored stable
 regions
 appear
  in the diagram 6a where $A\geq 0$. The respective stable window can be seen in the profile of $A$ in
  6b.
 \vspace{0.3cm}

\begin{figure}[!h]
 \begin{center}
 \includegraphics[scale=0.8]{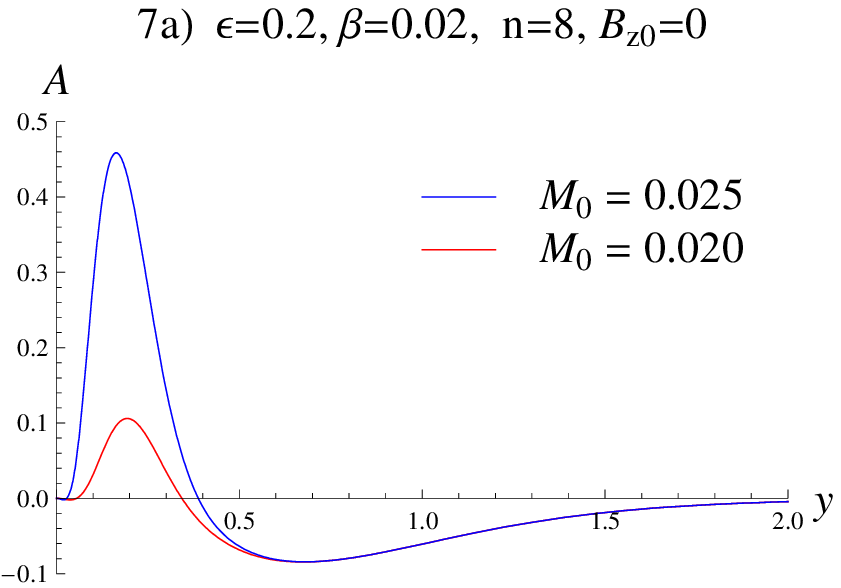}
 \includegraphics[scale=0.8]{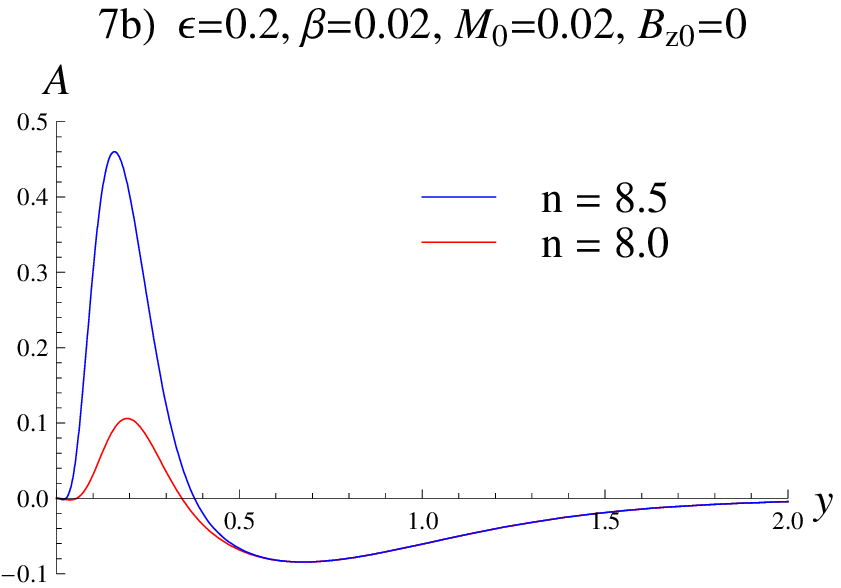}
 \includegraphics[scale=0.8]{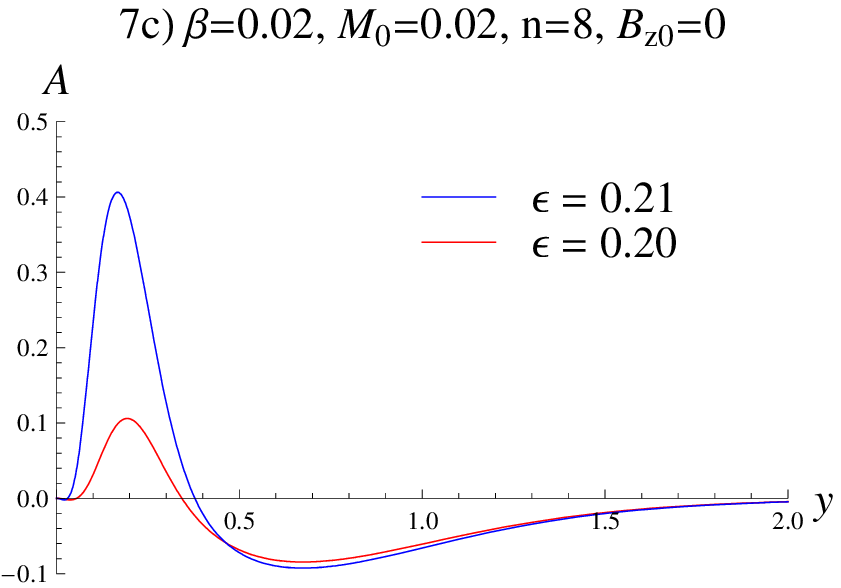}
 \includegraphics[scale=0.8]{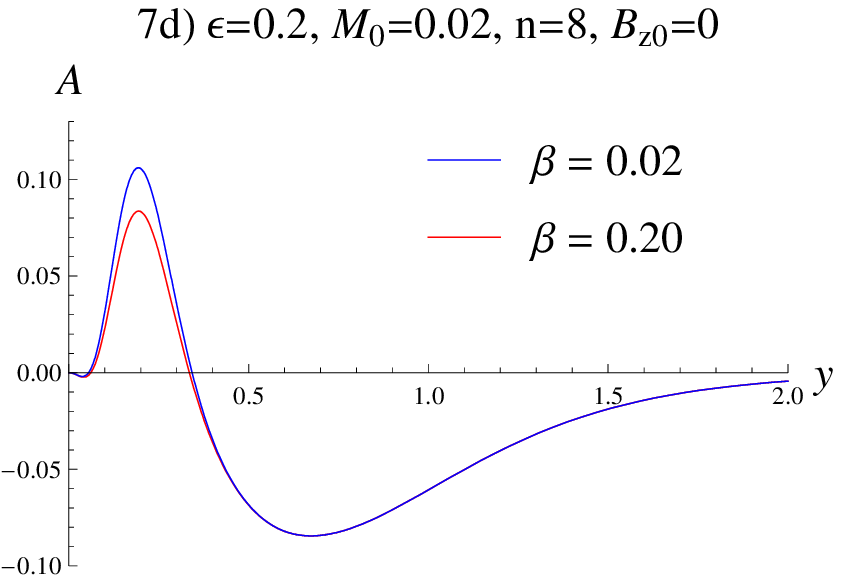}
 \caption{Impact of the flow (7a), flow   shear (7b), cat-eyes size
 (7c) and thermal pressure (7d)
 in connection with a variation of the parameters $M_0$,
 $n$ and $\epsilon$, and $\beta$, respectively, on the flow caused stable window
  associated with  $A\geq 0$ for the equilibrium of Fig. 3a.}
 \label{fig:7a}
 \end{center}
 \end{figure}

 \noindent
 Fig. \ref{fig:7a}: \ Impact of the flow (7a), flow   shear (7b), cat-eyes size
 (7c) and thermal pressure (7d)
 in connection with a variation of the parameters $M_0$,
 $n$ and $\epsilon$, and $\beta$, respectively, on the flow caused stable window
  associated with  $A\geq 0$ for the equilibrium of Fig. 3a.
  \vspace{0.3cm}

\begin{figure}[!h]
 \begin{center}
 \includegraphics[scale=0.8]{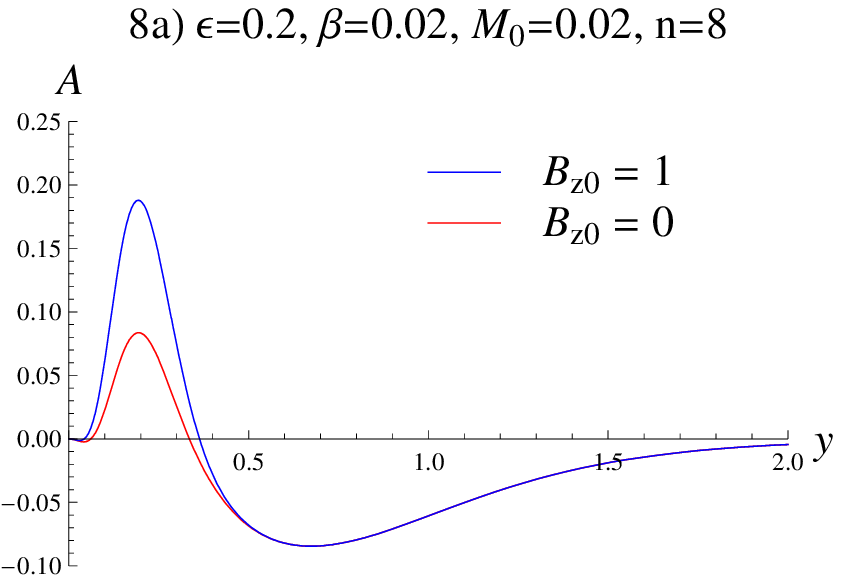}
 \includegraphics[scale=0.8]{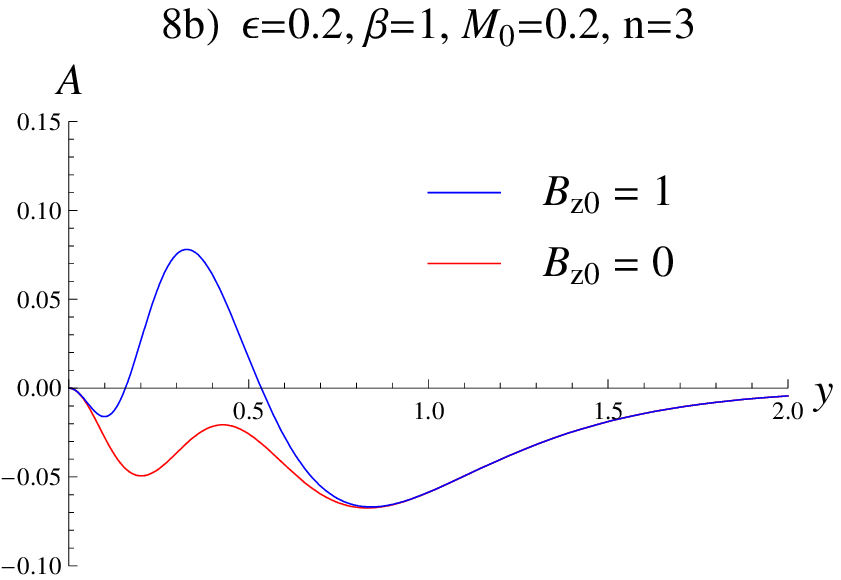}
 \caption{{\em Combined stabilization effect of flow and $B_{z0}$}:
 The curve 8a indicates a stabilizing synergism of $B_{z0}$  and  flow  for the equilibrium
 of Fig. 3a. A stronger synergism of this kind is shown  in Fig.
 8b pertaining to  a two-dimensional Harris-type equilibrium.}
 \label{fig:8a}
 \end{center}
 \end{figure}

  \noindent
 Fig. \ref{fig:8a}: \ {\em Combined stabilization effect of flow and $B_{z0}$}:
 The curve 8a indicates a stabilizing synergism of $B_{z0}$  and  flow  for the equilibrium
 of Fig. 3a. A stronger synergism of this kind is shown  in Fig.
 8b pertaining to  a two-dimensional Harris-type equilibrium.

 \newpage

 \begin{center}
 {\large \bf  List of Figures}
 \end{center}
  \vspace{-0.8cm}

\begin{thebibliography}{99}
\bibitem{qua} The term quasistatic has the meaning that the
               velocity term  is neglected in the
               momentum equation but   is kept in Ohm's law.
\bibitem{ind} Since the condition is sufficient when not satisfied ($A<0$), it becomes indecisive.
\end{thebibliography}
 \end{document}